\documentclass[12pt,tightenlines,floats,aps,prd,twoside,onecolumn,english,
	superscriptaddress,showkeys,preprintnumbers,nofootinbib,showpacs,
	amssymb,amsmath,amsfonts]{revtex4}

\usepackage{babel}
\usepackage{amsmath}
\usepackage{amsfonts}
\usepackage{amssymb}
\usepackage{dsfont}
\usepackage{graphicx}
\usepackage[usenames]{color}

\newcommand{\half}{\frac{1}{2}}
\newcommand{\halfi}{\frac{i}{2}}

\begin{document}

\preprint{IGC-09/3-4}

\setcounter{page}{1}

\title{Interaction of the Barbero--Immirzi Field with Matter and Pseudo-Scalar Perturbations}

\author{Simone Mercuri}
\email{mercuri@gravity.psu.edu}

\author{Victor Taveras}
\email{victor@gravity.psu.edu}

\affiliation{Institute for Gravitation and the Cosmos, The Pennsylvania State University,\\ Physics Department, Whitmore Lab, University Park, PA 16802, USA}

\date{August 31, 2009}

\begin{abstract}
In Loop Quantum Gravity the classical point of departure is the Einstein-Hilbert action
modified by the addition of the so-called Holst term. Classically, this term does not affect the equations of motion, but it induces a well-known quantization ambiguity in the quantum theory, parametrized by the Barbero--Immirzi parameter. Recently, it has been suggested to promote the Barbero--Immirzi parameter to a field. The resulting theory, obtainable starting from the usual Holst action, is General Relativity coupled to a pseudo-scalar field. However, this theory turns out to have an unconventional kinetic term for the BI field and a rather unnatural coupling with fermions.

The main goal of this work is twofold: Firstly, to propose a further generalization of the
Holst action, which yields a theory of gravity and matter with a more natural coupling to 
the Barbero--Immirzi field; secondly, to study the possible implications for cosmology correlated to the existence of this new pseudo-scalar field.
\end{abstract}



\maketitle

\section{Introduction}\label{sec:Intro}

Loop Quantum Gravity \cite{AshLew04,Rov04,Thi07} is a non-perturbative and mathematically rigorous formulation of a quantum theory of gravity. It is the result of the Dirac quantization procedure \cite{Dir64} applied to the Ashtekar--Barbero (AB) canonical constraints \cite{Ash86-87,Ash87-88,Bar95} of General Relativity (GR), which are classically equivalent to the usual constraints of canonical tetrad gravity in the temporal gauge. The equivalence of the two formulations relies on the fact that they essentially describe the same classical system in two different sets of fundamental variables, related by a one-parameter canonical transformation. According to Rovelli and Thiemann \cite{RovThi98}, this canonical transformation cannot be unitarily implemented in the quantum theory, so that the quantization procedure necessarily generates a one-parameter family of unitarily inequivalent representations of the quantum commutation relations. As a result, the spectra of the quantum geometrical observables are not uniquely determined, being affected by the presence of a constant, known as Barbero--Immirzi (BI) parameter. Specifically, the area spectrum is
\begin{equation}
A_{\gamma}=8\pi\gamma\ell_{P}^2\sum_k\sqrt{j_k\left(j_k+1\right)}\,,
\end{equation}
where $\gamma$ is the BI parameter, while $\ell_{P}$ is the Planck length. The numerical value of the BI parameter can be determined by studying the modes of isolated horizons of non-rotating black holes \cite{AshBaeCor98,AshBaeKra00}, but its physical origin is still an argument of discussions.

Recently, the idea that the BI parameter has a topological origin has shed light on the nature of this ambiguity, reinforcing the analogy with the so-called $\theta$-angle of Yang--Mills gauge theories as initially suggested by Gambini, Obregon and Pullin \cite{GamObrPul99}, and lately supported by other works \cite{Mer08,DatKauSen09,Mer09p1}.

It is worth remarking that in pure gravity this interpretation is not completely convincing, because the Holst action, which is the Lagrangian counterpart of the AB canonical formulation of gravity, does not contain any topological term. In fact, the so called Holst modification \cite{Hol96},\footnote{The signature throughout the paper is $+,-,-,-$, we set $\hbar=c=1$ and $\kappa=8\pi G$.}
\begin{equation}
S_{\rm Hol}\left[e,\omega\right]=\frac{1}{2\kappa\gamma}\int e_a\wedge e_b\wedge R^{ab}\,,
\end{equation} 
where $e^a$ is the gravitational field 1-form and $R^{a b}$ is the Riemann curvature 2-form defined as $R^{a b}=d\omega^{a b}+\omega^{a}_{\ c}\wedge\omega^{c b}$ ($\omega^{a b}$ being the Ricci spin connection 1-form), does not have the properties ascribable to a topological density, rather it is an on-(half)shell identically vanishing term. Nevertheless, the Holst framework can be further generalized and a true topological term can be naturally introduced in the action \cite{DatKauSen09,Mer09p1,CalMer09}, providing new interesting insights into the physical origin of the BI parameter \cite{Mer09p1}.

In order to clarify this point, let us summarize the main motivations to consider 
a generalization of the Holst framework. Generally, the coupling of fermion fields 
to first order Palatini gravity has a non-trivial effect on the geometry of 
spacetime: it generates a non-vanishing torsion tensor proportional to the spinor 
axial current. This fact has an interesting implication if we wish to describe the 
gravitational sector of the theory by using the Holst action, instead of the usual 
Hilbert--Palatini one. It turns out that in fact, in this more general case, the 
Holst modification no longer vanishes on-shell, consequently, the effective action 
differs from that of the Einstein--Cartan theory \cite{FreMinTak05,PerRov06,Mer06,Mer06p,BojDas08}. 
Specifically, the new effective action depends explicitly on the BI parameter, which, consequently, acquires a classical meaning \cite{PerRov06}. But, interestingly enough, the resulting modification diverges as soon as we set $\gamma=\pm\,i$, which respectively correspond to the self and anti-self dual formulation of gravity \cite{Ash86-87}. This induces one to look for a different formulation of the Holst action with matter \cite{Mer06}, satisfying the requirement that it exactly reduces to the Ashtekar-Romano-Tate action for $\gamma=\pm\,i$ \cite{AshRomTat89}.

In fact, the Holst term has a fermionic counter-term. In other words, it is possible to modify the Dirac as well as the gravitational action in such a way that the effective action exactly corresponds to the usual Einstein--Cartan one \cite{Mer06,Mer06p}. Basically, the two modifications sum up reconstructing the so-called Nieh--Yan topological density \cite{NieYan82}. This intriguing result was later confirmed in the framework of supergravities by Kaul \cite{Kau08}, who demonstrated that to preserve supersymmetry, the Holst modification of the gravitational sector has to be counterbalanced by a specific modification of the fermionic sector, which exactly corresponds to the one previously argued in \cite{Mer06} for the ordinary theory.

These results strongly suggest that the Nieh--Yan density plays an important role in gravity. In particular, its role seems to reflect that of the Chern--Pontryagin densities in Yang--Mills gauge theories \cite{Wei96,Pec98,Ash91}. Moreover, as advertised before, in this extended framework, the BI parameter can in fact be interpreted as a topological ambiguity analogous to the $\theta$-angle of Yang--Mills gauge theories \cite{Mer08,DatKauSen09,Mer09p1}. Specifically, such an ambiguity can be correlated to a specific large gauge sector of tetrad gravity in the temporal gauge \cite{Mer09p1}. This can also clarify why the appearance of the BI parameter is an unavoidable feature of the quantum theory of gravity based on the AB canonical formulation of GR.

So, we claim that a natural generalization of the Holst framework can be easily obtained by adding to the usual Hilbert--Palatini action, the Nieh--Yan density, i.e. \cite{DatKauSen09,CalMer09,Mer09,Mer09p1}
\begin{equation}\label{generalized action}
S\left[e,\omega\right]=-\frac{1}{2\kappa}\int e_a\wedge e_b\wedge\star R^{ab}+\frac{\beta}{2\kappa}\int\left(T^a\wedge T_a-e_a\wedge e_b\wedge R^{ab}\right)\,,
\end{equation}
where for later convenience we defined $\beta=-\frac{1}{\gamma}$. Above, $T^a$ is the torsion 2-form defined as $T^a=de^a+\omega^{a}_{\ b}\wedge e^b$, while the symbol ``$\star$'' denotes the Hodge dual operator. It is worth remarking that the above action is classically equivalent to the usual Hilbert--Palatini action, the Nieh--Yan being reducible to a total divergence, i.e.
\begin{equation}
\int\left(T^a\wedge T_a-e_a\wedge e_b\wedge R^{ab}\right)=\int d\left(e_a\wedge T^a\right)\,.
\end{equation}
Furthermore, action (\ref{generalized action}) can be straightforwardly extended to torsional spacetimes, in particular, it is dynamically equivalent to the unmodified Einstein--Cartan action, as can be easily demonstrated by considering spinor matter fields minimally coupled to gravity \cite{DatKauSen09}.

An intriguing possibility for a further non-trivial generalization of the theory is to promote the BI parameter to be a field. Some interesting results have been obtained performing this generalization in the Holst framework \cite{TavYun08} (see also \cite{GomKra09}).\footnote{Interestingly enough, the same model was considered a long time ago by Castellani, D'Auria and Fr\`e \cite{CasAurFre91}, who in a completely different framework, mainly suggested by String Theory, proposed the idea of considering a field interacting with gravity through the Holst modification. It is worth remarking that this proposal was precedent to the papers by Barbero and Immirzi: in this sense it has not any relation with the BI field considered in the recent works \cite{TavYun08} and \cite{GomKra09}.} 

Recently, a strong motivation to promote the BI parameter to be a field has been proposed by one of us \cite{Mer09}, clarifying, simultaneously, that the most natural starting point to study the dynamics of the BI field is, in fact, the following action
\begin{equation}\label{fund action}
S\left[e,\omega,\beta\right]=-\frac{1}{2\kappa}\int e_a\wedge e_b\wedge\star R^{ab}+\frac{1}{2\kappa}\int\beta(x)\left(T^a\wedge T_a-e_a\wedge e_b\wedge R^{ab}\right)\,,
\end{equation}
where the BI field couples to gravity via the Nieh--Yan term, instead of the Holst term. Essentially, the naturalness of the interaction between gravity and the BI field through the Nieh--Yan invariant is a consequence of the gravitational contribution to the chiral anomaly. It results that, in fact, the gravitational chiral anomaly in torsional spacetimes contains a divergent term proportional to the Nieh--Yan invariant. The necessity to reabsorb the divergence represents the motivation to promote the BI parameter to be a field \cite{Mer09}. This also clarifies the role played by the Nieh--Yan invariant in gravity, especially in relation with possible torsion sources as, e.g., spinor fields. In this respect, we recall that starting from the Holst action and promoting the BI parameter to be a field, it is $\varphi=\sinh\beta$ to play, in the effective action, the role of a scalar field rather than $\beta$ itself. Moreover, considering the presence of fermions, the Holst framework generates an unnatural coupling between the BI field and the fermionic matter fields \cite{GomKra09}, motivating, initially, the choice of a different and more natural starting point to describe the dynamics of the system in accordance with \cite{CalMer09}.

We note that, following a common procedure, the BI parameter can be associated with the expectation value of the field $\beta$ \cite{Mer09}. As a consequence its value can be correlated to some topological ambiguity through a dynamical Peccei--Quinn-like mechanism as, in fact, argued in \cite{Mer09}. 

In this paper, we clarify many aspects about the effective dynamics of the BI field, providing also some hints to study possible physical effects it could produce in a cosmological scenario. Specifically, the organization of the paper is the following: In Section \ref{sec II} we study the effective dynamics of the BI field, considering as a starting point the action (\ref{fund action}). In Section \ref{sec III} we discuss possible physical effects produced by the interaction of the BI field with matter. In particular, we calculate the low energy effective theory, extracting the mass of the BI field and the possible purely quantum contribution to the effective dynamics. In Section \ref{sec IV}, we argue that the BI field can solve the strong {\it CP} problem through the Peccei--Quinn dynamical mechanism. In fact, the effective action obtained in Section \ref{sec III} suggests to identify the BI field with the axion. The scale of the interaction and, as a consequence, the mass of the particle are both fixed by the theory. The large value of the coupling constant could be unnatural, motivating the argument of Section \ref{sec V}, where besides the BI field we consider also the presence of the axion. The coexistence of these two pseudo-scalar degrees of freedom is particularly interesting as they interact with matter in a linear combination. Physically, only one particular linear combination can acquire an anomaly induced mass, so that we expect one massive and one massless physical particle as a result of the coexistence of the BI field and the axion. The massless particle interacting with bosonic matter can produce interesting effects, in particular we consider the perturbations it can produce on the cosmic microwave background (CMB), and digress on the known experimental limit on the magnitude of this effect. 

In the Appendix \ref{App A}, we further generalize the theory by relaxing the scale of the interaction between the BI field and gravity, introducing a dimensional parameter $M$. Surprisingly enough, this generalization does not change the scale of the effective interaction between the BI field and matter, which is determined by the theory itself. Appendix \ref{App B} is devoted to a brief description of the effective dynamics of the BI field coupled to the original Holst modification, so that the outcomes of the different models can be easily compared. A discussion of the proposed model concludes the paper.

\section{Effective Dynamics of Nieh--Yan Modified Gravity with Fermions}\label{sec II}

In order to study the dynamics of the BI field $\beta$, we consider a further generalization of the action (\ref{fund action}), by introducing fermion fields, which will play an essential role in our argumentation. So, let us couple spinor fields to gravity through the usual minimal prescription, i.e.\footnote{For notations see \cite{Mer09p2}.}
\begin{align}\label{fund action with fermions}\nonumber
S\left[e,\omega,\beta,\psi,\overline{\psi}\right]=&-\frac{1}{2\kappa}\int e_a\wedge e_b\wedge\star R^{ab}+\frac{1}{2\kappa}\int\beta(x)\left(T^a\wedge T_a-e_a\wedge e_b\wedge R^{ab}\right)
\\
&+\halfi\int\star e_a\wedge\left(\overline{\psi}\gamma^a D\psi-\overline{D\psi}\gamma^a\psi+\frac{i}{2}\,m e^a\overline{\psi}\psi\right)\,.
\end{align}
The covariant derivatives are defined as
\begin{subequations}\label{cov derivatives}
\begin{align}
D\psi=d\psi-\frac{i}{4}\,\omega^{a b}\Sigma_{a b}\psi\,,
\\
\overline{D\psi}=d\overline{\psi}+\frac{i}{4}\,\overline{\psi}\Sigma_{a b}\omega^{a b}\,,
\end{align}
\end{subequations}
where $\Sigma^{a b}=\frac{i}{2}\left[\gamma^a,\gamma^b\right]$ are the generators of the Lorentz group.

Now by varying the action with respect to the dynamical fields, we can extract the equations of motion. First, let us calculate the equation resulting from the variation with respect to the connection 1-form $\omega^{ab}$, i.e.
\begin{equation}
	\label{eq:7}
	\frac{1}{4\kappa}\,\epsilon_{abcd}D(e^a\wedge e^b)+\frac{1}{8}\,\star e_a\bar{\psi}\{\gamma^a,\Sigma_{cd}\}\psi+\frac{1}{2\kappa}\,e_c\wedge e_d\wedge d\beta=0\,,
\end{equation}
which, by using the formula $\left\{\gamma^a,\Sigma^{b c}\right\}=2\epsilon^{abc}_{\ \ \ d}\gamma^5\gamma^d$, can be rewritten as
\begin{equation}
	\label{eq:7b}
	\frac{1}{4\kappa}\,\epsilon_{abcd}(T^a\wedge e^b-e^a\wedge T^b)-\frac{1}{4}\,\epsilon_{abcd}\star e^a J_{(A)}^b+\frac{1}{2\kappa}\,e_c\wedge e_d\wedge d\beta=0\,,	
\end{equation}
where $J_{(A)}^b=\bar{\psi}\gamma^b\gamma^5\psi$ is the spinor axial current. This equation can be further reduced with some algebra; we have, 
\begin{equation}\label{structure equation}
T^a\wedge e^b-e^a\wedge T^b-\star e^{[a}\left(\kappa J_{(A)}^{b]}-2\eta^{b] c}\partial_c\beta\right)=0\,.
\end{equation}
The variation of the action in Eq. (\ref{fund action with fermions}) with respect to $\beta$, $\overline{\psi}$, $\psi$ and $e^a$ respectively yields,
\begin{subequations}
\begin{align}
e_a\wedge e_b\wedge R^{ab}-T^a\wedge T_a&=0\,,\label{NY equation}
\\
\star e_a\wedge\left(i\gamma^a D\psi-\frac{m}{4}\,e^a\psi\right)&=0\,,
\\
\star e_a\wedge\left(i\overline{D\psi}\gamma^a+\frac{m}{4}\,e^{a}\overline{\psi}\right)&=0\,,
\\\nonumber
\frac{1}{\kappa}\,e_b\wedge\star R^{ab}-\frac{1}{\kappa}\,d\beta\wedge T^a
+\frac{i}{4}\star\left(e^a\wedge e_b\right)
\\
\wedge\left(\overline{\psi}\gamma^b D\psi-\overline{D\psi}\gamma^b\psi\right)-\star e^a m\overline{\psi}\psi&=0\,.
\end{align}
\end{subequations}

The set of the equations of motion is complicated, nevertheless interesting consequences can be extracted from the effective dynamics. First, let us re-express the connection 1-form $\omega^{ab}$ as function of the other dynamical fields. In this respect, we recall that the connection $\omega^{ab}$ has to satisfy the second Cartan structure equation
\begin{equation}\label{inhomogeneus}
de^a+\omega^{a}_{\ b}\wedge e^b=T^a\,,
\end{equation}
which, being a linear equation, admits a natural decomposition of the connection. Specifically, recalling that the contorsion 1-form $K^{a b}$ captures the part of the gravitational connection that depends on torsion, a natural decomposition of the connection 1-form $\omega^{a b}$ is
\begin{equation}
	\label{eq:14}
	\omega^{ab}={}^o\!\omega^{ab}(e)+K^{ab}\,.
\end{equation}
$^o\!\omega^{ab}(e)$ is the usual Ricci spin connection and satisfies the homogeneous structure equation, namely
\begin{equation}
	\label{eq:15}
	de^a+{}^o\!\omega^a{}_b\wedge e^b=0\,,
\end{equation}
while the contorsion 1-form is related to the torsion 2-form as follows
\begin{equation}
	\label{eq:16}
	K^a{}_b\wedge e^b=T^a\,,
\end{equation}
so that the full connection 1-form in \eqref{eq:14} satisfies the inhomogeneous structure equation (\ref{inhomogeneus}).

As a first step we extract the expression of the torsion 2-form from Eq. (\ref{structure equation}): we easily obtain
\begin{equation}\label{torsion}
T^a=-\frac{1}{2}\,\star\left[e^a\wedge e_b\left(\kappa J_{(A)}^{b}-2\eta^{b f}\partial_f\beta\right)\right]=-\frac{1}{4}\,\epsilon^{a}_{\ b c d}\left(\kappa J_{(A)}^{b}-2\eta^{b f}\partial_f\beta\right)e^c\wedge e^d\,.
\end{equation}
It is important to note that, in order to preserve the standard transformation properties of the torsion tensor under the Lorentz group, the field $\beta$ has to be a pseudo-scalar \cite{CalMer09,Mer09}. In other words, the geometrical content of the theory suggests the pseudo-scalar nature of the BI field $\beta$, which is a consequence of the peculiar interaction with the Nieh--Yan density and is not assumed \emph{a priori}. The explicit expression for torsion in (\ref{torsion}) corresponds to the following expression for the contortion 1-form,
\begin{equation}\label{contortion}
K^{ab}=\frac{1}{4}\,\epsilon^{ab}_{\ \ c d}e^c\left(\kappa J_{(A)}^{d}-2\eta^{d f}\partial_f\beta\right)\,.
\end{equation}
Hence, the full connection 1-form satisfying the inhomogeneous second Cartan structure equation is
\begin{equation}\label{solution connection}
\omega^{ab}={}^o\omega^{ab}(e)+\frac{1}{4}\,\epsilon^{ab}_{\ \ c d}e^c\left(\kappa J_{(A)}^{d}-2\eta^{d f}\partial_f\beta\right)\,.
\end{equation}

Now, by substituting the solution (\ref{solution connection}) into the other equations of motion we can study the effective dynamics. It is particularly interesting to note that by equation (\ref{NY equation}) we obtain
\begin{equation}
\star d\!\star\! d\beta=\frac{\kappa}{2}\,\star\!d\!\star\! J_{(A)}=m\kappa\,\overline{\psi}\gamma^5\psi\,,
\end{equation}
where we have used $\star d\!\star\! J_{(A)}=2m\overline{\psi}\gamma^5\psi$, $J_{(A)}=J_{(A)}^a e_a$ being the axial current 1-form and $d V$ the natural volume element. This equation establishes a dynamical relation between the BI field and the composite pseudo-scalar $\overline{\psi}\gamma^5\psi$, implying that the BI field has to be a pseudo-scalar too, as previously suggested by geometrical arguments.

It is worth noting that the same effective equations can be obtained by pulling back the action (\ref{fund action with fermions}) on the solution of the structure equation (\ref{solution connection}), obtaining the effective action, namely
\begin{align}\label{effective action}\nonumber
	S_{\rm eff}=&-\frac{1}{2\kappa}\int e_a\wedge e_b\wedge\star {}^o\!R^{ab}+\halfi\int \star e_a\wedge\left(\overline{\psi}\gamma^a \,{}^o\!D\psi-\overline{{}^o\!D\psi}\gamma^a\psi+\frac{i}{2}e^a m\overline{\psi}\psi\right)
\\
&+\frac{3}{16}\kappa\int\star J_{(A)}\wedge J_{(A)}+\frac{3}{4\kappa}\int\star d\beta\wedge d\beta-\frac{3}{4}\int\star J_{(A)}\wedge d\beta\,,
\end{align}
and varying it with respect to the other dynamical fields. It is worth remarking that the  action (\ref{effective action}) contains only torsionless objects, in particular the symbol ``$\,{}^o\,$'' denotes that the 2-form ${}^o\!R^{ab}$ is the curvature associated with the Ricci spin connection ${}^o\!\omega^{ab}$ and, analogously, the covariant derivatives action on spinors are defined as in (\ref{cov derivatives}), where the full connection is replaced by ${}^o\!\omega^{ab}$. The non-vanishing torsion tensor contributes to the kinetic term of the field $\beta$ and generates the four-fermion Fermi-like term as well as the interaction between the field $\beta$ and the spinor axial current. The pseudo-scalar nature of the BI field prevents the theory from any possible parity violation, in contrast with what was argued elsewhere in the literature \cite{GomKra09}.

Finally, in order to reabsorb the constant factor in front of the kinetic term for the BI field, we can define the new dimensional field $\phi$ by rescaling the original adimensional field $\beta$ in a suitable way, i.e.
\begin{equation}
\phi:=\sqrt{\frac{3}{2\kappa}}\,\beta\,,
\end{equation}
so that the last two terms of the effective action can be rewritten as follows
\begin{equation}\label{interaction}
\frac{3}{4\kappa}\int\star d\beta\wedge d\beta-\frac{3}{4}\int\star J_{(A)}\wedge d\beta=\frac{1}{2}\int\star d\phi\wedge d\phi-\frac{\sqrt{6\kappa}}{4}\int\star J_{(A)}\wedge d\phi\,.
\end{equation}
The last term above contains interesting dynamical information about the interaction of the rescaled BI field and matter: the study of this interaction will be the focus of the next section.

\section{Interaction with matter}\label{sec III}

This Section is dedicated to study the properties of the interactions between the BI field and ordinary matter. For the sake of clarity, we divided the discussion into two parts: in the first one, we consider the classical effective theory, extracting some physical properties of the BI field; in the second part, we take into account also some purely quantum effects, correlated with the existence of the chiral anomaly.

\subsection{Classical effective theory}

The interaction between the rescaled BI field $\phi$ and spinor matter through the last term in (\ref{interaction}) is particularly interesting in view of the description of the dynamics of the dynamical BI field. In particular, it is worth noting that the interaction vanishes in the case of massless fermions, which preserve chirality. The case of massive fermions, though, is more interesting from a physical perspective. In this case, in fact, the interaction term can be reabsorbed by transforming the spinor fields in a suitable way. The non-trivial dynamical content of this interaction reflects in the mass term by introducing a modification in the low energy theory.

Specifically, the interaction contained in the last term of (\ref{interaction}) can be reabsorbed by operating the following transformation:
\begin{subequations}\label{chiral rotation}
\begin{align}
\psi\,\to\,\psi^{\prime}&=e^{-i\frac{\sqrt{6\kappa}}{4}\phi\gamma^5}\psi\,,
\\
\overline{\psi}\,\to\,\overline{\psi}^{\prime}&=\overline{\psi}e^{-i\frac{\sqrt{6\kappa}}{4}\phi\gamma^5}\,,
\end{align}
\end{subequations}
which modifies the mass term in such a way that the effective action becomes:
\begin{align}\label{effective action mass}\nonumber
S_{\rm eff}=&-\frac{1}{2\kappa}\int e_a\wedge e_b\wedge\star {}^o\!R^{ab}
+\frac{1}{2}\int\star\, d\phi\wedge d\phi+\frac{3}{16}\kappa\int\star J_{(A)}\wedge J_{(A)}
\\
&+\halfi\int \star e_a\wedge\left(\overline{\psi}\gamma^a \,{}^o\!D\psi-\overline{{}^o\!D\psi}\gamma^a\psi+\frac{i}{2}\,e^a m\overline{\psi}e^{-i\frac{\sqrt{6\kappa}}{2}\phi\gamma^5}\psi\right)\,.
\end{align}
Now, in order to extract a low energy effective action for the field $\phi$ and the neutral pion $\pi^0$, let us consider the modified mass term in the Lagrangian for the first two generations of quarks, i.e. the up and down quark, respectively denoted by $u$ and $d$. We have 
\begin{align}
L_{\rm mass}=-m_u\overline{u}e^{-i\frac{\phi}{F_{\phi}}\gamma^5}u-m_d\overline{d}e^{-i\frac{\phi}{F_{\phi}}\gamma^5}d\,.
\end{align}
Now by using the following formulas \cite{Wei96}
\begin{subequations}
\begin{align}
\overline{u}u\to -v\cos\left(\frac{\pi^0}{F_{\pi}}\right)\,,&\qquad\overline{d}d\to -v\cos\left(\frac{\pi^0}{F_{\pi}}\right)\,,
\\
\overline{u}\gamma^5 u\to -iv\sin\left(\frac{\pi^0}{F_{\pi}}\right)\,,&\qquad\overline{d}\gamma^5 d\to iv\sin\left(\frac{\pi^0}{F_{\pi}}\right)\,,
\\\label{third line}
i\overline{u}\gamma^a\gamma^5 u\to \frac{1}{2}F_{\pi}\partial^a\pi^0+\dots\,,&\qquad i\overline{d}\gamma^a\gamma^5 d\to \frac{1}{2}F_{\pi}\partial^a\pi^0+\dots\,,
\end{align}
\end{subequations}
where $F_{\pi}\simeq 93{\rm MeV}$ is the pion decay constant, measured in the decay $\pi^+\to\mu^+ +\nu_{\mu}$ and $v=\langle\overline{u}u\rangle=\langle\overline{d}d\rangle$ is a constant, we can rewrite the mass term in the Lagrangian as:
\begin{equation}
L_{\rm mass}=m_u v\cos\left(\frac{\pi^0}{F_{\pi}}-\frac{\phi}{F_{\phi}}\right)+m_d v\cos\left(\frac{\pi^0}{F_{\pi}}+\frac{\phi}{F_{\phi}}\right)\,,
\end{equation}
where for convenience we defined $F_{\phi}=\frac{2}{\sqrt{6\kappa}}\simeq 2.0\times 10^{18}{\rm GeV}$, which determines the scale of the interaction between the field $\phi$ and matter. It is important to note that this constant is not a free parameter, but it is fixed by the theory.\footnote{In the Appendix \ref{App A} we tried to introduce a free parameter in the topological sector of the initial action in order to relax the value of the scale of the interaction, but surprisingly enough, the theory is not sensitive to this modification and the scale of the interaction cannot be naively relaxed.}

Interestingly enough, by making the replacements in (\ref{third line}), the four fermion term in the effective action (\ref{effective action mass}) generates a modification of the kinetic term of the effective low energy pion Lagrangian, namely:
\begin{equation}
\frac{3}{16}\,\kappa J^a J_a=-\frac{1}{16}\,\frac{F_{\pi}^2}{F_{\phi}^2}\,\partial_a\pi^0\partial^a\pi^0\,.
\end{equation}
Since $F_{\phi}\gg F_{\pi}$ we can safely neglect this modification, so that the low energy theory is, essentially, unaffected by the four fermion interaction.

In order to extract informations about the low energy limit, we are mainly interested in the quadratic part of the Lagrangian for the pion and BI field. In particular, the modified mass term represents a non-trivial potential which deserves to be studied, because it could produce an interesting dynamics. But for the scope of this work we only consider the quadratic part, postponing a deeper investigation for a following work \cite{MerTavYun09}. So, expanding the Lagrangian up to the second order in the fields, we obtain the following result: 
\begin{equation}
L_{\rm mass}\simeq -\frac{1}{2}\begin{array}{cc}\left(\pi^0\right.&\left.\phi\right)\\
\ &\ \end{array}M^2\left(\begin{array}{c}\pi^0 \\ \phi\end{array}\right)\,,
\end{equation}
where the $2\times 2$ matrix $M^2$ is given by
\begin{equation}
M^2=\left(\begin{array}{cc}
\left(m_u+m_d\right)\frac{v}{F_{\pi}^2}&\left(-m_u+m_d\right)\frac{v}{F_{\pi}F_{\phi}}
\\
\left(-m_u+m_d\right)\frac{v}{F_{\pi}F_{\phi}}&\left(m_u+m_d\right)\frac{v}{F_{\phi}^2}
\end{array}\right)\,.
\end{equation}
Now, considering that $F_{\phi}\gg F_{\pi}$, one of the eigenvalues of the above mass matrix corresponds to the mass of the pion, namely $m_{\pi}^2=\left(m_u+m_d\right)\frac{v}{F_{\pi}^2}$, while the other is the mass of the BI field $\phi$. By a simple calculation, we obtain 
\begin{equation}
m_{\phi}=\frac{F_{\pi}}{F_{\phi}}m_{\pi}\frac{\sqrt{m_u m_d}}{m_u+m_d}\simeq 3.0\times 10^{-12}{\rm eV}\,.
\end{equation}
So, this procedure has allowed to evaluate the mass of the BI field, which is completely determined by the theory. The small mass of the field $\phi$ makes this particle undetectable in accelerator experiments. Moreover, by calculating the cross section of the process of decaying into two photons, it results that its lifetime is longer than the present age of the Universe; so that the BI field can travel for a cosmological distance before decaying.

\subsection{Quantum effects}

Let us now make a detour from the classical effective theory, considering also possible quantum effects. So far we have only considered fermions interacting with the gravitational field. As a further generalization, we consider here that the spinor matter fields interact also with a generic gauge field, described by an $SU(N)$ connection, $A=A^I\lambda^I$, where $\lambda^I$ are the $N^2-1$ hermitian generators of the group. The presence of this new gauge field, minimally coupled to gravity and matter, does not affect the solution of the second Cartan structure equation calculated in the previous section (i.e. Eq. (\ref{solution connection})). So that, by defining the new covariant derivative acting on spinors as $\mathcal{D}={}^o\!D+i A$, the effective action becomes
\begin{align}\label{effective action III}\nonumber
S_{\rm eff}=&-\frac{1}{2\kappa}\int e_a\wedge e_b\wedge\star{}^o\!R^{ab}-\frac{1}{2}\int{\rm tr}\star F\wedge F+\frac{1}{2}\int\star d\phi\wedge d\phi
\\
&+\halfi\int\star e_a\wedge\left(\overline{\psi}\gamma^a  \mathcal{D}\psi-\overline{\mathcal{D}\psi}\gamma^a\psi+\frac{i}{2}\,e^a m\overline{\psi}\psi\right)
\\\nonumber
&+\frac{3}{16}\kappa\int\star J_{(A)}\wedge J_{(A)}-\frac{\sqrt{6\kappa}}{4}\int\star J_{(A)}\wedge d\phi\,,
\end{align}
where $F=dA+\left[A,A\right]$ is the curvature 2-form associated to the gauge field $A$.

As above, the last term in the effective action can be reabsorbed by a chiral transformation of the spinor fields. But as soon as we take into account possible quantum effects, the existence of the quantum chiral anomaly introduces a purely quantum contribution to the effective action besides the modification affecting the mass term. To be more precise, let us suppose that we quantize the theory by using the path-integral method. The fermionic measure is not invariant under a chiral rotation \cite{Fuj79-80}, generating a contribution to the divergence of the axial current proportional to the Pontryagin densities associated with the gauge fields contained in the theory. Specifically, we have that \cite{Ber96}:
\begin{equation}
d\star J_{(A)}=2m\overline{\psi}\gamma^5\psi d V+\frac{1}{8\pi^2}\,{\rm tr}F\wedge F+\frac{1}{8\pi^2}\,R^{ab}\wedge R_{ab}\,.
\end{equation} 
So, by considering the last term in the effective action (\ref{effective action III}) and the expression of the chiral anomaly above, we can expect an interaction between the BI and the gauge fields. In fact, such an interaction can be predicted by using a quite general argument, completely neglecting the presence of fermionic fields. Indeed, once assumed that the BI parameter is actually a field, the resulting effective action is that of a pseudo-scalar field decoupled from gravity. The effective action is symmetric under a shift of the new field, thus showing the existence of an additional symmetry belonging to the group $U_{(A)}(1)$. This symmetry cannot remain unbroken in the quantum regime and an anomaly must appear. The anomalous contribution to the effective action due to the axial rescaling of the field $\beta$ corresponds to that of a (pseudo-)Nambu-Goldstone boson correlated to an existing broken symmetry (see \cite{Pec98,Wei96}). In this framework, $\beta(x)$ plays a role strictly analogous to that of  the axion, also suggesting that it can be used to solve the so-called \emph{strong {\it CP} problem}, as will be argued in the next section. 

Let us complete this Section further motivating the expected full semi-classical action. As we showed before, the interaction term between $\phi$ and the spinor fields can be easily eliminated by the chiral rotation (\ref{chiral rotation}). But taking into account the purely quantum contribution coming from the chiral rotation of the fermionic measure, a new non-trivial interaction between $\phi$ and the gauge fields appears in the effective action, namely
\begin{align}\label{effective semi-classical action III}\nonumber
S_{\rm eff}=&-\frac{1}{2\kappa}\int e_a\wedge e_b\wedge\star{}^o\!R^{ab}-\frac{1}{2}\int{\rm tr}\star F\wedge F+\frac{1}{2}\int\star d\phi\wedge d\phi
\\
&+\halfi\int\star e_a\wedge\left[\overline{\psi}\gamma^a  \mathcal{D}\psi-\overline{\mathcal{D}\psi}\gamma^a\psi+\frac{i}{2}\,e^a m\overline{\psi}e^{-i\frac{\sqrt{6\kappa}}{2}\phi\gamma^5}\psi\right]
\\\nonumber
&+\frac{1}{8F_{\phi}^2}\int\star J_{(A)}\wedge J_{(A)}-\frac{1}{8\pi^2}\int\frac{\phi}{2F_{\phi}}{\rm tr} F\wedge F-\frac{1}{8\pi^2}\int\frac{\phi}{2F_{\phi}}\, R^{ab}\wedge R_{ab}\,.
\end{align}

On the one hand, it is interesting to note that the sector of the effective action describing the dynamics of the gravitational and pseudo-scalar field reproduces the so-called
Chern-Simons Modified gravity \cite{JacPi03,AleYun08,AleFinYun08,SopYun09}. On the other hand, as stressed before, the explicit expression of the effective action above suggests to use the BI field to implement the Peccei--Quinn mechanism in order to solve the strong {\it CP} problem: that will be the focus of the next section.

\section{Solving the strong {\it CP} problem with the BI field}\label{sec IV}

Let us begin this Section by introducing the strong {\it CP} problem. As is well known, the vacuum angle of QCD, $\theta$, is a possible source of {\it CP} violation. In the standard model, the $U(1)\times SU(2)$ symmetry breaking is another source for {\it CP} violation, in particular the non-hermitian quark mass matrix, $M$, introduces a {\it CP} violating factor proportional to ${\rm Arg}\det M$. Experimentally, one expects a violation of the {\it CP} symmetry proportional to $\tilde{\theta}=\theta+{\rm Arg}\det M$, but, observing the neutron electric dipole moment, an expected and unnatural small upper limit of the order of $10^{-10}$ can be fixed for the parameter $\tilde{\theta}$. This small value implies an extremely precise fine tuning between the two parameters entering in $\tilde{\theta}$. This compensation is rather unnatural because the two parameters $\theta$ and ${\rm Arg}\det M$ are completely independent; namely $\theta$ is motivated by the non-trivial topological structure of the group $SU(3)$ of QCD, while $M$ is correlated to the breaking of the $SU(2)\times U(1)$ symmetry. This is known as strong {\it CP} problem.

The Peccei--Quinn mechanism \cite{PecQui77} is a dynamical solution of the strong {\it CP} problem. It is mainly based on the assumption that a new global chiral $U(1)_A$ symmetry is present in the Lagrangian of the standard model, usually this additional symmetry is denoted as $U(1)_{PQ}$. It is easy to demonstrate that if the chiral Peccei--Quinn symmetry is an exact symmetry of the Standard Model, the strong {\it CP} problem can be easily solved. Interestingly enough, even though the $U(1)_{PQ}$ is spontaneously broken, it is still possible to solve the strong {\it CP} problem through a dynamical mechanism, namely the PQ mechanism \cite{PecQui77}. 

Essentially, the dynamical PQ mechanism works as follows. The postulated $U(1)_{PQ}$ symmetry cannot remain exact at the quantum level: it is, in fact, spontaneously broken by the presence of the chiral anomaly. The spontaneous breaking of this symmetry reflects in the presence of a pseudo-scalar Nambu--Goldstone boson, called axion, the vacuum expectation value of which is non-vanishing and driven by a non-trivial anomaly-induced potential to the value which compensates the {\it CP} violating parameter, $\tilde{\theta}$. In this Section we argue that the role of the axion can be played directly by the BI field, with the advantage that it has a correlation with the gravitational theory through the Nieh--Yan topological term. If this argument works properly, an interesting bridge can be constructed between particle physics and gravity, through a topological sector mainly motivated by the classical framework of Loop Quantum Gravity.

Let us enter in the details by reconsidering the effective action (\ref{effective semi-classical action III}). In particular, let us specialize the action to describe a particular coupled system of matter and gravity: the bosonic gauge connection 1-form $A$ is valued on $SU(3)$ and the coupling constant $g$, entering in the definition of the full covariant derivative $\mathcal{D}$, is the strong charge. For the sake of simplicity, let us focus our attention only on the matter content of the effective action. Moreover, in order to make contact with the standard framework, let us also suppose that the spacetime is described by a solution of the Einstein equations, such that the term $R_{ab}\wedge R^{ab}$ identically vanishes (this class of solutions contains many  physically relevant spacetimes as, e.g., Friedmann-Robertson-Walker and Schwarzschild). Finally, the effective action we want to consider is
\begin{align}\label{effective action IV}\nonumber
S_{\rm eff}&= S_{\rm Grav}[e]+S_{\rm Strong}[A]+S_{\rm Ferm}[\psi_f,\overline{\psi}_f,e,A,\phi]+S[\phi]
\\
&+\frac{1}{8F_{\phi}^2}\int\sum_f\star J_{(A)f}\wedge J_{(A)f}
+\frac{g^2}{8\pi^2}\int\left(\tilde{\theta}-\frac{\phi}{2F_{\phi}}\right){\rm tr} G\wedge G\,,
\end{align}
where $G$ is the curvature 2-form associated with the strong interaction. Above, we considered the {\it CP}-violating term proportional to the parameter $\tilde{\theta}$ as well as the presence of different quarks flavors denoted by the symbol $f$. 

The interaction between the field $\phi$ and the gluons in the last term of the action above is a consequence of the quantum contribution of the chiral anomaly, as explained in the previous Section. It is worth noting that the {\it CP} violating $\tilde{\theta}$-term combines with the $\phi$ field, so that the possible observables of the theory now depend on the combination $\tilde{\theta}-\frac{\phi}{2F_{\phi}}$. Since experiments exclude a {\it CP} violation of this kind, according to the Peccei--Quinn mechanism, we can conclude that the effective potential driving the dynamics of the BI field has to be even in $\tilde{\theta}-\frac{\phi}{2F_{\phi}}$, so that it has a stationary point in 
\begin{equation}
\tilde{\theta}-\frac{\phi}{2F_{\phi}}=0\,.
\end{equation}

So, it seems that the BI field can be a possible candidate to play the role of the axion, regarding a dynamical solution of the strong {\it CP} problem through the Peccei--Quinn mechanism. The scale of the interaction between $\phi$ and matter is fixed by the theory to be of the order of $F_{\phi}$, which corresponds to a extremely light and weakly interacting axion. This is a precise prediction of the theory and deserves to be seriously taken into consideration. If cosmological and astrophysical data excluded the identification of the BI field with the PQ axion, then a different possible scenario could be taken into consideration, namely the existence of two pseudo-scalar fields, which can be respectively associated to the BI field and to the standard axion. The consequences of this hypothesis are described in the next Section.

\section{Pseudo Scalar Perturbations}\label{sec V}

According to what said at the end of the previous Section, let us consider the possible coexistence of the BI field and the standard axion, correlated with the dynamical solution of the strong {\it CP} problem.

We expect that the axion and the BI field combine and naturally interact via the chiral anomaly with the gauge bosons in a linear combination. Considering also the presence of the electromagnetic field $F$, schematically we have \cite{PosRitSko08}:
\begin{equation}
\mathcal{L}_{\rm int}=\left(\frac{\phi}{2g_{\phi}}+\frac{a}{2g_{a}}\right){\rm tr}G\wedge G+\left(\frac{\phi}{2f_{\phi}}+\frac{a}{2f_{a}}\right)F\wedge F+\left(\frac{\phi}{2r_{\phi}}+\frac{a}{2r_{a}}\right)R^{a b}\wedge R_{a b}\,,
\end{equation}
where $g_{\phi},f_{\phi},r_{\phi}$ and $g_{a},f_{a},r_{a}$ determine respectively the scales of the interactions of the fields $\phi$ and $a$ with the gluon, electromagnetic and gravitational fields.\footnote{The scale of the interaction of the field $\phi$ are determined by the theory as stressed more than once. Specifically, they are correlated to the constant $F_{\phi}$ defined above and rescaled by the square of the coupling constant, e.g. $f_{\phi}^{-1}=\frac{\alpha}{2\pi}F_{\phi}^{-1}$, where $\alpha$ is the fine structure constant.} 

As is well known, the axion acquires an induced mass term through the chiral anomaly. The mass of the axion directly depends on the energy scale at which the Peccei--Quinn chiral symmetry breaks down. Interestingly enough, the mechanism that induces a mass is peculiar and only one linear combination of the two pseudo-scalar fields acquires a mass \cite{AnsUra82}. At an effective level, this fact implies that besides the usual QCD term for the massive physical axion, one has a massless additional pseudo-scalar field, $\Phi$, which interacts with the electromagnetic as well as the gravitational field as follows:
\begin{equation}
\mathcal{L}_{\rm int}=\mathcal{L}_{\rm axion}+\frac{\Phi}{2f_{\Phi}}F\wedge F+\frac{\Phi}{2r_{\Phi}}R^{a b}\wedge R_{a b}\,,
\end{equation}
where $f^{-1}_{\Phi}$ and $r^{-1}_{\Phi}$ denote the scale of the respective interactions.

The presence of the coupling with photons induces a rotation of the polarization, $\varepsilon$, of an electromagnetic wave, according to the following expression \cite{HarSik92}
\begin{equation}\label{polarization perturbation}
\Delta\varepsilon=\frac{\Delta\Phi}{f_{\Phi}}\,,
\end{equation}
where $\Delta\Phi$ is the spacetime variation of the massless pseudo-scalar field. 

So the existence of the BI field, motivated by the necessity of reabsorbing a divergence in the chiral anomaly in torsional spacetime as recently argued in \cite{Mer09}, combined with the pseudo-scalar field associated to the additional Peccei--Quinn symmetry \cite{PecQui77}, leads to the existence of a massless pseudo-scalar field $\Phi$. Interestingly enough, from a cosmological point of view, the existence of this massless state super-weakly interacting with photons has interesting effects on the polarization of CMB. In particular, by studying the polarization anisotropies of the observed spectra of CMB, it is possible to put a stringent lower bound on the scale parameter $f_{\Phi}$, as recently demonstrated by Pospelov, Ritz, and Skordis \cite{PosRitSko08}, who have also proved that this method can efficiently probe new models containing such new pseudo-scalar fields. They have found that $f_{\Phi}>2.4\times 10^{14}{\rm GeV}$, fixing a constraint on massless pseudo-scalars more stringent of at least two order of magnitude with respect to previously existing limits.

\section{Discussion}\label{Discussion}

In this paper, according to previous works \cite{TavYun08,GomKra09,CalMer09,Mer09}, we proposed to promote the BI parameter to a field. Initially, this idea was motivated by the hope of associating the constant value of the BI parameter to the expectation value of a field through a dynamical mechanism. But, simultaneously, the presence of this new field in the action allows to solve another problem correlated with the chiral anomaly on a torsional spacetime \cite{Mer09}. In particular, according to a result of Chand\'ia and Zanelli \cite{ChaZan97}, the chiral anomaly diverges on spacetimes with torsion, the divergence being correlated to the Nieh--Yan contribution to the anomaly itself. This divergence can be reabsorbed in the definition of the BI field, so its presence naturally solves this problem. It is worth remarking that an analogous redefinition could work even if we did not consider the possibility that $\beta$ is actually a field. In other words, we could imagine to reabsorb the divergence in the physical BI parameter, considering the $\beta$ appearing in the gravitational action as a ``bare'' vacuum parameter. But this redefinition would involve a shift of the parameter and to work properly requires a sort of ``fine tuning''. So the solution with the BI field seems to be preferable and based on a shift symmetry of the theory immediately recognizable looking at action (\ref{fund action with fermions}) or (\ref{effective action}).

The dynamical equations show that this new field is a pseudo-scalar, suggesting two interesting perspectives. In fact, studying the effective action and, in particular, the peculiar interaction of the BI field with ordinary matter, one immediately realizes that the BI field can be used to implement the Peccei--Quinn mechanism to solve the strong {\it CP} problem. Interestingly enough, the scale of the interaction between the BI field and matter in this model is fixed by the theory to such a value that the anomaly induced mass for the BI field is determined by the theory and corresponds to an extremely light and weakly interacting axion-like particle. 

By taking seriously this suggestion of the theory, we are left with two possibilities. One is that the PQ axion, which dynamically solves the strong {\it CP} problem, is well described by this model, mainly motivated by the necessity of reabsorbing a divergence in the chiral anomaly. The other possibility is that the BI field and the axion are, in fact, different fields, existing simultaneously. In such a framework, it is possible that the physical pseudo-scalar states associated to the two physical particles are a linear superposition of the BI and the axion fields. This second possibility is interesting from a cosmological point of view, because if the two fields interact with the gluons as well as photons through a linear superposition, only one of the two physical pseudo-scalar degrees of freedom can acquire an anomaly-induced mass, the other remaining massless. As a consequence, the massless field, interacting with photons generates a rotation of the polarization angle of electromagnetic waves and this effect can be probed by studying the polarization anisotropies of CMB.

In general, any massless pseudo-scalar field super-weakly coupled to photons generates such an effect, which, compared to the available data on the B-mode, allows to fix a limit on the strength of the pseudo-scalar coupling to photons. Many new high-energy physical theories contains or predict two or more light pseudo-scalar fields, which can generate such an effect. Our model with the BI field represents a possible new theoretical framework in which a new pseudo-scalar particle is present and can, in fact, possibly encompass the physics generating a rotation of the polarization angle of CMB in a fairly standard way.

\acknowledgments
We would like to thank Gianluca Calgani, Nicolas Yunes for discussions
and especially Abhay Ashtekar for having originally suggested 
this project. This research was supported in part by NSF grant PHY0854743, The George A.
and Margaret M. Downsbrough Endowment and the Eberly research funds of
Penn State. VT acknowledges support from the Alfred P. Sloan Foundation, 
and the Eberly College of Science.

\appendix

\section{On the Role of Scale Parameter}\label{App A}

Here we consider a further generalization of the action \eqref{fund action with fermions} for Nieh--Yan gravity by introducing a parameter $M$ with the dimensions of an energy,\footnote{It is worth remarking that now the BI field $\tilde{\beta}$ has the dimensions of an energy as well.}
\begin{align}\label{eq:action with M}\nonumber
S\left[e,\omega,\beta,\psi,\overline{\psi}\right]=&-\frac{1}{2\kappa}\int e_a\wedge e_b\wedge\star R^{ab}-M\int\tilde{\beta}(x)\left(e_a\wedge e_b\wedge R^{ab}-T^a\wedge T_a\right)
\\
&+\halfi\int\star e_a\wedge\left(\overline{\psi}\gamma^a D\psi-\overline{D\psi}\gamma^a\psi+\frac{i}{2}\,m e^a\overline{\psi}\psi\right)\,.
\end{align}
The introduction of the parameter $M$ is extremely natural and generalizes the action in Eq. (\ref{fund action with fermions}) by relaxing the scale of the interaction between matter and the BI field $\beta$. In fact, we saw that the BI field 
couples to the Nieh--Yan density via a term of the form
\begin{equation}
	\frac{1}{2\kappa}\int\beta(x)(e_a\wedge e_b\wedge R^{ab}-T^a\wedge T_a)\,,
\end{equation}
and the coupling scale is uniquely determined by the factor $\frac{1}{2\kappa}$ ($\kappa=8\pi G$ in natural units, i.e. $c=\hbar=1$, where $G$ is the Newton's constant).

One can then proceed with this action along the lines in the main body of the paper and obtain the effective action,
\begin{align}\label{effective action with M}\nonumber
	S_{\rm eff}=-\frac{1}{2\kappa}\int e_a\wedge e_b\wedge\star{}^o\!R^{ab}+\halfi\int \star e_a\wedge\left(\overline{\psi}\gamma^a \,{}^o\!D\psi-\overline{{}^o\!D\psi}\gamma^a\psi+\frac{i}{2}e^a m\overline{\psi}\psi\right)
\\
+\frac{3}{16}\kappa\int\star J_{(A)}\wedge J_{(A)}+3\kappa M^2\int \star d\tilde{\beta}\wedge d\tilde{\beta}-\frac{3}{2}\kappa M \int\star J_{(A)}\wedge d\tilde{\beta}\,.
\end{align}

Then the last two terms in the effective action suggest a field redefinition of the form
\begin{equation}
	\phi=\sqrt{6\kappa}M\tilde{\beta}\,,
\end{equation}
upon which the effective action takes the form
\begin{align}\nonumber
	S_{\rm eff}=-\frac{1}{2\kappa}\int e_a\wedge e_b\wedge\star {}^o\!R^{ab}+\halfi\int \star e_a\wedge\left(\overline{\psi}\gamma^a \,{}^oD\psi-\overline{{}^oD\psi}\gamma^a\psi+\frac{i}{2}e^a m\overline{\psi}\psi\right)
\\
+\frac{3}{16}\kappa\int\star J_{(A)}\wedge J_{(A)}+\half\int\star d\phi\wedge d\phi-\frac{\sqrt{6\kappa}}{4}\int \star J_{(A)}\wedge d\phi \,.
\end{align}
Surprisingly, one obtains a coupling between $\phi$ and fermionic matter which is \emph{independent 
of the parameter $M$}. Thus, even though we have tried to allow an arbitrary energy scale for the coupling of the BI field, it turns out that the scale of the interaction is fixed by the theory to the Planck energy.

\section{On BI field in Holst Gravity}\label{App B}

\subsection{Effective Dynamics}

In \cite{TavYun08} a different model for the BI field was considered. The same model was already studied much earlier in \cite{CasAurFre91}, before the Holst,
Barbero, and Immirzi proposals became popular in the Quantum Gravity community. The model in \cite{CasAurFre91} was most likely motivated by String Theory considerations and led to a theory of General Relativity coupled to a pseudo-scalar field $\beta$, which remarkably was not the consequence of the ``scalarization'' of the BI parameter.

Here we compare the dynamical features of the BI field as resulting from Holst gravity with that presented in this paper and explain why the present model is more appealing and, to a great extent, more natural from a practical perspective. Note the changes in convention:
our field $\beta$ corresponds to the field $-\bar{\gamma}$ of \cite{TavYun08},
lowercase internal indexes $abc$ correspond to uppercase 
indexes $IJK$ of \cite{TavYun08}, our contorsion tensor is denoted $K^{ab}$ rather than $C^{IJ}$, and finally our metric signature is $+,-,-,-$ whereas $-+++$ was used in \cite{TavYun08}. Moreover, we stress that in \cite{TavYun08} the Riemann 2-form was denoted by $F^{a b}$, using the symbol $R^{a b}$ for the torsion-free curvature; here we used the usual symbol $R^{a b}$ for the full curvature and ${}^o\!R^{ab}$ for the torsion-free Riemann 2-form. 

Neglecting matter, the action considered in \cite{TavYun08} is
\begin{equation}
	\label{eq:a1}
	S=-\frac{1}{4\kappa}\int \epsilon_{abcd}e^a\wedge e^b\wedge R^{cd}-\frac{1}{2\kappa}\int \beta e_a\wedge e_b\wedge R^{ab}\,.
\end{equation}
At once one notices that the second term in \eqref{eq:a1} is the original Holst modification, but the $T^a\wedge T_a$ contribution, which makes up the other part of the 
so-called Nieh--Yan invariant, is missing. This is significant because \eqref{eq:a1} yields a theory with torsion as shown in \cite{TavYun08} and the $T^a\wedge T_a$ term affects the dynamical outcomes when $\beta$ is non-constant.

One can then vary the action with respect to $\omega^{ab}$ and solve the Cartan
structure equation for the torsion 2-form thus obtaining equation (15) in \cite{TavYun08},
\begin{equation}
	\label{eq:a2}
T^a=\frac{1}{2}\frac{1}{1+\beta^2}\left(\epsilon^a{}_{bcd}\partial^d\beta+\beta\delta^a_{[b}\partial_{c]}\beta\right)e^b\wedge e^c
\end{equation}
and the corresponding contorsion 1-form
\begin{equation}
	\label{eq:a3}
	K^{ab} =-\frac{1}{2}\frac{1}{1+\beta^2}\left(\epsilon^{ab}{}_{cd} e^c\partial^d\beta+2\beta e^{[a}\partial^{b]}\beta\right)\,.
\end{equation}
There are two main differences between these formulas obtained in \cite{TavYun08} and the
corresponding expressions obtained here in \eqref{torsion} and \eqref{contortion}. Firstly,
the vector trace component of the torsion tensor \eqref{torsion} vanishes, whereas it does not 
for \eqref{eq:a2}. Secondly, the expressions \eqref{eq:a2} and
\eqref{eq:a3} contain a $\beta$ dependent prefactor of $\frac{1}{1+\beta^2}$ which
complicated the theory. The effective action then yields GR coupled to the scalar field
$\varphi=\sinh\beta$ rather than $\beta$, affecting also the coupling with fermions that becomes quite unnatural \cite{GomKra09}.

\subsection{Effective Dynamics with Fermions}

Alternatively, one can also take the theory described by \eqref{eq:a1}, still without the complete Nieh--Yan term, and non-minimally couple fermions to it using the non-minimal 
coupling introduced in \cite{Mer06} and \cite{Mer08}. Then one gets a more complicated 
theory than the one presented in this work. Its action is
\begin{align}
	\label{eq:a4}
S(e,\omega,\psi,\bar{\psi},\beta)=&-\frac{1}{4\kappa}\int \epsilon_{a b c d}e^a\wedge e^b\wedge R^{c d}
-\frac{1}{2\kappa}\int\beta e_a\wedge e_b\wedge R^{ab}
\nonumber
\\
&+\halfi\int\star e_a\wedge\left[\overline{\psi} \gamma^a(1-i\beta\gamma^5)D\psi-\overline{D\psi}(1-i\beta\gamma^5)\gamma^a\psi\right]\,. 
\end{align}
The structure equation can be calculated by varying with respect to $\omega$, i.e.
\begin{equation}
	\label{eq:a5}
\left(\half\,\epsilon^{ab}_{\ \ cd}+\beta\delta_c^{[a}\delta_d^{b]}\right)d^{(\omega)}(e^a\wedge e^b)-\left(\half\,\epsilon^{ab}_{\ \ cd}+\beta\delta_c^{[a}\delta_d^{b]}\right)\kappa\star e^c J^d+e^a\wedge e^b\wedge d\beta=0\,.
\end{equation}
One can then solve \eqref{eq:a5} for the torsion 2-form thus obtaining,
\begin{equation}
	\label{eq:a10}
	T^a=-\frac{1}{4}\,\epsilon^a_{\ bcd}\left(\kappa J^b-\frac{2}{1+\beta^2}\,\partial^b\beta\right)e^c\wedge e^d + \frac{1}{2}\frac{\beta}{1+\beta^2}\,e^a\wedge d\beta\,,
\end{equation}
and the corresponding contorsion 1-form
\begin{equation}
	\label{eq:a11}
	K^{ab} =\frac{1}{4}\,\epsilon^{ab}{}_{cd} e^c\left(\kappa J^d-\frac{2}{1+\beta^2}\,\partial^d\beta\right)+\frac{\beta}{1+\beta^2}\,e^{[a}\partial^{b]}\beta\,.
\end{equation}
This is similar to the result obtained in the main body of this work, but for the presence of the factor $\frac{1}{1+\beta^2}$ and the fact that the torsion tensor (\ref{torsion}) has a vanishing trace component. The theory in the main body of this work has no $\beta$-dependent multiplicative factors nor do such
$\beta$-dependent multiplicative factors show up in the effective theory either. Whereas, for the theory
\eqref{eq:a4} the effective action is more complicated
\begin{align}
	\label{eq:a12}
S_{\rm eff}=&-\frac{1}{4\kappa}\int \epsilon_{abcd}e^a\wedge e^b \wedge {}^o\!R^{cd}+\halfi\int \star e_a\wedge\left(\overline{\psi}\gamma^a \,{}^o\!D\psi-\overline{{}^o\!D\psi}\gamma^a\psi\right) \nonumber 
\\
&+\frac{3}{16}\kappa\int dV\eta_{ab}J^a J^b
-2\int dV J^a\partial_a\beta+\frac{3}{4\kappa}\int dV\frac{1}{1+\beta^2}\,\eta^{ab}\partial_a\beta\partial_b\beta\,.
\end{align}

Finally, we note that in order to have a standard kinetic term for the scalar field, a change of variable is necessary, specifically we have to introduce the field $\varphi=\sinh\beta$, but this inevitably complicates the interaction with fermionic matter. Summarily, the simplicity of \eqref{effective action}, obtained from the theory described by action
\eqref{fund action with fermions}, suggests that considering the full Nieh--Yan term leads to more appealing and natural results, when the interaction with fermions is taken into account.


\begin{thebibliography}{99}

\bibitem{AshLew04}
A. Ashtekar and J. Lewandowski, \emph{Class. Quant. Grav.} \textbf{21}, R53 (2004).

\bibitem{Rov04}
C. Rovelli, \emph{Quantum Gravity} (Cambridge University Press, Cambridge, 2004).

\bibitem{Thi07}
T. Thiemann, \emph{Modern Canonical Quantum General Relativity} (Cambridge University Press, Cambridge, 2007).

\bibitem{Dir64}
P.A.M. Dirac, \emph{Lectures on Quantum Mechanics} (Belfer Graduate School of Science, Yeshiva University, New York, 1964).

\bibitem{Ash86-87}
A. Ashtekar, \emph{Phys. Rev. Lett.} \textbf{57}, 2244 (1986); \emph{Phys. Rev.} \textbf{D36}, 1587 (1987).

\bibitem{Ash87-88}
A. Ashtekar, \emph{Mathematics and General Relativity} (American Mathematical Society, Providence, Rhode Island, 1987).

\bibitem{Bar95}
F. Barbero, \emph{Phys. Rev.} \textbf{D51}, 5498 (1995); \emph{Phys. Rev.} \textbf{D51}, 5507 (1995).

\bibitem{RovThi98}
C. Rovelli and T. Thiemann, \emph{Phys. Rev.} \textbf{D57}, 1009 (1998).

\bibitem{AshBaeCor98}
A. Ashtekar, J. Baez, A. Corichi, and K. Krasnov, \emph{Phys. Rev. Lett.} \textbf{80}, 904 (1998).

\bibitem{AshBaeKra00}
A. Ashtekar, J. Baez, and K. Krasnov, \emph{Adv. Theor. Math. Phys.} \textbf{4}, 1 (2000).

\bibitem{GamObrPul99}
R. Gambini, O. Obregon, and J. Pullin, \emph{Phys. Rev.} \textbf{D59}, 047505 (1999).

\bibitem{Mer08}
S. Mercuri, \emph{Phys. Rev.} \textbf{D77}, 024036 (2008).

\bibitem{DatKauSen09}
G. Date, R.K. Kaul, and S. Sengupta, \emph{Phys. Rev.} \textbf{D79}, 044008 (2009). 

\bibitem{Mer09p1}
S. Mercuri, \texttt{arXiv:0903.2270}.

\bibitem{Hol96}
S. Holst, \emph{Phys. Rev.} \textbf{D53}, 5966 (1996).

\bibitem{CalMer09}
G. Calcagni and S. Mercuri, \emph{Phys. Rev.} \textbf{D79}, 084004 (2009).

\bibitem{FreMinTak05}
L. Freidel, D. Minic, and T. Takeuchi, \emph{Phys. Rev.} \textbf{D72}, 104002 (2005).

\bibitem{PerRov06}
A. Perez and C. Rovelli, \emph{Phys. Rev.} \textbf{D73}, 044013 (2006).

\bibitem{Mer06}
S. Mercuri, \emph{Phys. Rev.} \textbf{D73}, 084016 (2006).

\bibitem{Mer06p}
S. Mercuri, Proceedings of the Eleventh Marcel Grossmann Meeting on General Relativity, Berlin (Germany), July 23-29, 2006, eds. H. Kleinert, R.T. Jantzen, and R. Ruffini (World Scientific, Singapore, 2008).

\bibitem{BojDas08}
M. Bojowald and R. Das, \emph{Class. Quant. Grav.} \textbf{25}, 195006 (2008). 

\bibitem{AshRomTat89}
A. Ashtekar, J.D. Romano, and R.S. Tate, \emph{Phys. Rev.} \textbf{D40}, 2572 (1989).

\bibitem{NieYan82}
H.T. Nieh and M.L. Yan, \emph{J. Math. Phys.} \textbf{23}, 373 (1982).

\bibitem{Kau08}
R.K. Kaul, \emph{Phys. Rev.} \textbf{D77}, 045030 (2008).

\bibitem{Wei96}
S. Weimberg, \emph{The quantum theory of fields} (Cambridge University Press, Cambridge 1996), Vol. 2.

\bibitem{Pec98}
R.D. Peccei, \texttt{arXiv:hep-ph/9807516}.

\bibitem{Ash91}
A. Ashtekar (in collaboration with R.S. Tate), \emph{Lectures on non-perturbative canonical garvity} (World Scientific, Singapore, 1991).

\bibitem{Mer09}
S. Mercuri, \emph{Phys. Rev. Lett.} \textbf{103}, 081302 (2009).

\bibitem{TavYun08}
V. Taveras and N. Yunes, \emph{Phys. Rev.} \textbf{D78}, 064070 (2008).

\bibitem{GomKra09}
A. Torres-Gomez and K. Krasnov, \emph{Phys. Rev.} \textbf{D79}, 104014 (2009).

\bibitem{CasAurFre91}
L. Castellani, R. D'Auria, and P. Fre, \emph{Supergravity and Superstrings} (World Scientific, Singapore, 1991).

\bibitem{Mer09p2}
S. Mercuri, \emph{Proc. Sci.} ISFTG (2009) 016.

\bibitem{MerTavYun09}
S. Mercuri, V. Taveras, and N. Yunes, (unpublished).

\bibitem{Fuj79-80}
K. Fujikawa, \emph{Phys. Rev. Lett.} \textbf{42}, 1195 (1979); \emph{Phys. Rev.} \textbf{D21}, 2848 (1980).

\bibitem{Ber96}
R.A. Bertlmann, \emph{Anomalies in Quantum Field Theory} (Clarendon Press, Oxford, 1996).

\bibitem{JacPi03}
R. Jackiw and S.Y. Pi, \emph{Phys. Rev.} \textbf{D68}, 104012 (2003).

\bibitem{AleYun08}
S. Alexander and N. Yunes, \emph{Phys. Rev.} \textbf{D77}, 124040 (2008).

\bibitem{AleFinYun08}
S. Alexander, L.S. Finn, and N. Yunes, \emph{Phys. Rev.} \textbf{D78}, 066005 (2008).

\bibitem{SopYun09}
C.F. Sopuerta and N. Yunes, \emph{Phys. Rev.} \textbf{D80}, 064006 (2009).

\bibitem{PecQui77}
R.D. Peccei and H.R. Quinn, \emph{Phys. Rev. Lett.} \textbf{38}, 1440 (1977); \emph{Phys. Rev.} \textbf{D16}, 1791 (1977).

\bibitem{PosRitSko08} 
M. Pospelov, A. Ritz and C. Skordis, \texttt{arXiv:0808.0673}.

\bibitem{AnsUra82}
A.A. Anselm and N.G. Uraltsev, \emph{Phys. Lett.} \textbf{B114}, 39 (1982).

\bibitem{HarSik92}
D. Harari and P. Sikivie, \emph{Phys. Lett.} \textbf{B289}, 67 (1992).

\bibitem{ChaZan97}
O. Chand\'{i}a and J. Zanelli, \emph{Phys. Rev.} \textbf{D55}, 7580 (1997).

\end{thebibliography}
\end{document}